\documentstyle[twoside,fleqn,npb,epsfig]{article}   
\newcommand{\AmS}{{\protect\the\textfont2   
  A\kern-.1667em\lower.5ex\hbox{M}\kern-.125emS}}   
   
\title{ 
Initial conditions and evolution of off-diagonal distributions
}  
   
\author{ 
\underline{K.~Golec-Biernat}\address{H.~Niewodniczanski Institute of Nuclear Physics,  
ul.~Radzikowskiego 152, Krakow, Poland}, A.D.~Martin\address{Department of Physics,  
University of Durham, DH1 3LE, United Kingdom} 
and M.G.~Ryskin\address{St.~Petersburg  
Nuclear Physics Institute, Gatchina, St.~Petersburg, 188350, Russia}   
}  
   
\begin{document}   
   
\begin{abstract}    
We briefly discuss the problem of specifying initial conditions for  
evolution of off-diagonal (skewed) parton distributions. We present 
numerical results to show that evolution rapidly washes out differences of 
input.
\end{abstract}   
   
\maketitle   
   
\section{Introduction} 
Off-diagonal (or skewed) parton distributions provide important information 
about the nonperturbartive structure of the nucleon \cite{JI}. They can in principle be  
measured 
in such processes as deeply virtual Compton scattering, diffractive vector meson production 
or diffractive high-$p_T$ jet production. Just as for the ordinary (diagonal) parton  
distributions, 
the QCD evolution equations \cite{BELITSKY} play an important role in the determination 
of skewed parton  
distributions. As usual input distributions are required.  However, this is more complicated 
than in the diagonal case since the skewed parton  
distributions depend 
on additional  variable -- the asymmetry parameter $\xi \sim p-p^{\prime}$.  
We use the  symmetric formulation of Ji \cite{JI} in which the skewed 
parton distributions are given by functions $H (x, \xi)$  
with support $-1 \leq (x, \xi) \leq 1$, see \cite{MARTIN} for more details. Here 
we discuss various ways to specify the initial conditions and show how the differences in the  
input distributions disappear on evolution.  In particular 
 we illustrate the conclusion of \cite{MARTIN} that the skewed distributions 
$H (x, \xi)$, at small $x$ and $\xi$, are fixed by  the conventional diagonal partons. 
 
\section{Constraints imposed on $H(x,\xi)$} 
The distributions $H(x,\xi)$ have to fulfil several conditions with respect to 
the variable $\xi$. First,  the time-reversal invariance and hermiticity  
impose the condition \cite{JI} 
\begin{equation} 
H(x,\xi)\,=\,H(x,-\xi)\,. 
\label{eq:2} 
\end{equation} 
Thus $H(x,\xi)$ is an even function of $\xi$. 
% and from now we only consider $\xi\ge 0$. 
The second condition states that in the limit $\xi=0$ we recover the ordinary 
diagonal parton distributions 
\begin{equation} 
H(x,0)\,=\,H^{diag}(x)\,. 
\label{eq:3} 
\end{equation} 
The third condition is more complicated but has a simple origin,  
\cite{JI,RAD1}. The $N^{th}$ moment of $H(x,\xi)$ is a polynomial 
in $\xi$ of the order $N$ at most 
\begin{equation} 
\int_{-1}^1 dx\, x^{N-1}\, H(x,\xi)\,=\,\sum_{i=0}^{[N/2]}\;A_{N,i}\;\xi^{2i}\,, 
\label{eq:4} 
\end{equation} 
which also embodies condition (\ref{eq:2}).  Finally we impose continuity of $H(x,\xi)$ at  
the border $x=\pm\xi$ 
between two different physical regions, see \cite{MARTIN}.  
This ensures that the amplitude of a physical process, described by skewed distributions, is 
finite.  All these conditions have to be fulfilled in the construction of the input for evolution. 
They are, of course, 
conserved during the evolution. 
 
\section{Initial conditions for evolution} 
 
There are two equivalent ways to evolve $H(x, \xi, \mu^2)$ up in the  scale 
$\mu^2$. The first method, which is most appropriate at small $x, \xi$, uses the evolution of  
the Gegenbauer moments of $H(x,\xi)$ and 
the Shuvaev transform \cite{SHUV,SGMR} to find the final answer in  
the $x-$space \cite{MARTIN}. 
%This method is described in \cite{MARTIN}.  
In the second 
method the solutions are found after imposing 
initial conditions and numerically solving the evolution equations,  
directly in the $x-$space.  Here we supplement the studies of \cite{MARTIN}  
by presenting results from the second approach.  
 
Condition (\ref{eq:4}) is difficult to fulfil in order to specify initial distributions 
at a  certain scale $\mu_0^2$.  It may be facilitated 
by the use of the double distribution ${\cal{F}}(\tilde{x},\tilde{y})$ \cite{RAD1} 
\begin{equation} 
H(x,\xi) \;=\; \int_{\cal{R}} d\tilde{x} d\tilde{y} \; 
{\cal F}(\tilde{x},\tilde{y})\; \delta(x-(\tilde{x}+\xi \tilde{y}))\;, 
\label{eq:5} 
\end{equation} 
where ${\cal{R}}$ is the square $|\tilde{x}|+|\tilde{y}|\le 1$
This prescription introduces a nontrivial 
mixing between $x$ and $\xi$. 
Condition (\ref{eq:2}) is guaranteed if  
${\cal{F}}(\tilde{x},\tilde{y})={\cal{F}}(\tilde{x},-\tilde{y})$.  
We still have freedom to add to (\ref{eq:5}) a function $\mbox{\rm sign}(\xi)\, D(x/\xi)$,  
antisymmetric in $x/\xi$, contained entirely in the ERBL-like region $|x|<\xi$  
\cite{POLYAKOV}. 
 
The only problem left is to build in the ordinary diagonal distributions in the prescription 
(\ref{eq:5}).  To do this we take 
\begin{equation} 
{\cal F}(\tilde{x},\tilde{y})\;=\,h(\tilde{y})\,H^{diag}(\tilde{x})\; /\; 
\int_{-1+|\tilde{x}|}^{1-|\tilde{x}|} dy^\prime\,h(y^\prime)\,, 
\label{eq:6} 
\end{equation} 
where different choices of $h$ give, via (\ref{eq:5}), different initial conditions for $H (x,  
\xi)$.  Three choices, 
\begin{equation} 
h({\tilde{y}})\; =\; \cases{\delta(\tilde{y}) \cr  
(1-{\tilde{y}}^2\,)^{p+1} ~~~~~p=0,1 
\cr \sin(\pi\,\tilde{y}^2)}, 
\label{eq:7} 
\end{equation} 
are shown in Fig.~1, with $H^{diag} (x)$ given by GRV \cite{GRV} at  
$\mu_0^2=0.26~{\mbox{\rm GeV}}^2$.  The first choice gives simply the diagonal input  
$H^{diag}(x)$, independent of $\xi$.  The second, with $p=0 (1)$ for quarks (gluons),  
generates an input form similar to that obtained  from the model of \cite{MARTIN,SGMR}  
in which the Gegenbauer moments of $H(x,\xi)$  are $\xi$-independent. This property
is conserved by the evolution. 
The exact form 
of the double distribution in this case can be found in \cite{RAD2}.  The last choice was  
selected so as to give an oscillatory input behaviour in the ERBL-like region.  
 
\section{Discussion} 
 
Fig.~1 shows the quark non-singlet, quark singlet and gluon skewed distributions for the three  
input models evolved up in $\mu^2$ for $\xi = 0.03$. The results for larger values of $\xi$  
are qualitatively the same \cite{GMR}. 
 
We see that the form of $h(\tilde{y})$ has the most influence in the 
ERBL region. However even in this region evolution soon washes out the differences.  
Already by $\mu^2=100$~GeV$^2$ 
the three curves  are close to each other. In the DGLAP region, $|x|>\xi$, the  
curves are almost  identical while in the  ERBL region, $|x|<\xi$, they approach the  
asymptotic form for each particular skewed parton distribution. 
Such behaviour is especially important in view of the fact that one 
set of the curves is obtained from the pure diagonal input parton distributions.  
In this way we illustrate the result presented in \cite{MARTIN} that to a good accuracy  
at small $\xi$, the skewed distributions $H (x, \xi; \mu^2)$ are  
completely known in terms of conventional partons.   
Thus, to summarize,  
the nonperturbative information contained in the diagonal input parton distributions 
and particular features of the evolution equations for skewed parton distributions 
are sufficient for their determination. 

\medskip 
\noindent {\bf Acknowledgements} 
 
\medskip 
We thank Max Klein and Johannes Bl\"{u}mlein for their efficient organization of DIS99,  
and the Royal Society and the EU Fourth Framework Programme `Training and Mobility of 
Researchers', Network `QCD and the Deep Structure of Elementary Particles', contract 
FMRX-CT98-0194 (DG 12-MIHT) for support.

\newpage 
 
\begin{figure*} 
 \epsfig{figure=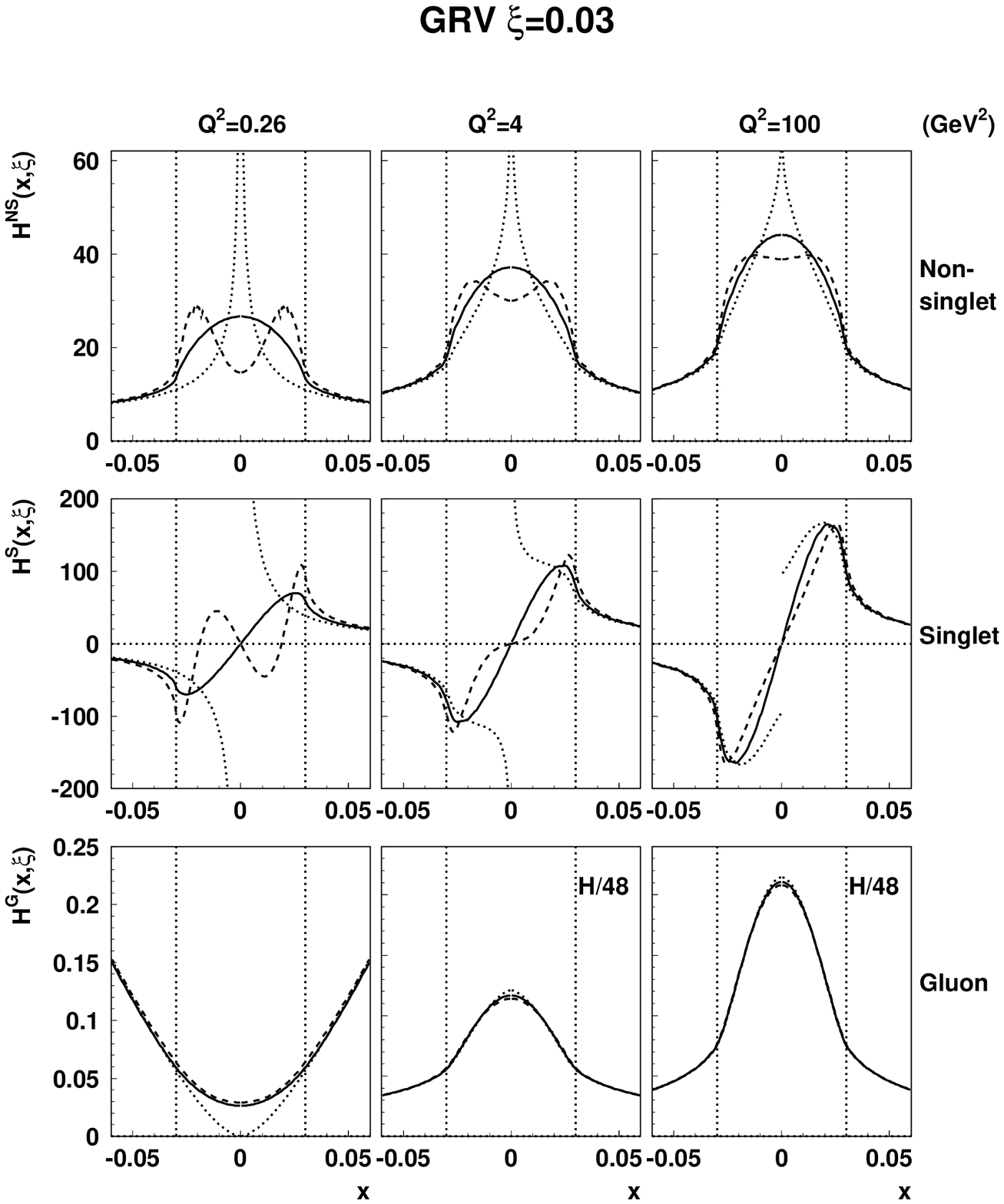,width=15cm} 
\vspace{-1.2cm} 
\caption{The evolution to $\mu^2 = Q^2=4$ and $100~{\mbox{\rm GeV}}^2$ of the quark  
non-singlet $H^{NS}$, quark singlet $H^S$ and the gluon distribution $H^G$ 
starting from different inputs at $\mu_0^2= 0.26~{\mbox{\rm GeV}}^2$ for $\xi =  
0.03$ (indicated by the dotted vertical lines $x=\pm \xi$).  The dotted curves correspond to  
evolution from diagonal input, the continuous curves to the model of [3] 
and the dashed curves to oscillatory input in the ERBL-like region.} 
\end{figure*} 
   
\end{document}